\documentstyle [12pt,aps,amsfonts] {revtex}
\input epsf
\newcommand{\beq}{\begin{equation}}
\newcommand{\eeq}{\end{equation}}
\newcommand{\beqs}{\begin{eqnarray}}
\newcommand{\eeqs}{\end{eqnarray}}
\newcommand{\lsim}{\mathrel{\raisebox{-.6ex}{$\stackrel{\textstyle<}{\sim}$}}}
\newcommand{\gsim}{\mathrel{\raisebox{-.6ex}{$\stackrel{\textstyle>}{\sim}$}}}

\begin{document}
\tighten
\draft

\baselineskip 6.0mm

\title{Effects of Gauge Interactions on Fermion Masses in Models with 
Fermion Wavefunctions Separated in Higher Dimensions} 

\vspace{8mm}

\author{
Shmuel Nussinov$^{(a,b)}$ \thanks{email: nussinov@post.tau.ac.il} \and
Robert Shrock$^{(b)}$ \thanks{email: robert.shrock@sunysb.edu}}

\vspace{6mm}

\address{(a) \ Sackler Faculty of Science \\
Tel Aviv University \\
Tel Aviv, Israel} 

\address{(b) \ C. N. Yang Institute for Theoretical Physics \\
State University of New York \\
Stony Brook, N. Y. 11794, USA}

\maketitle

\vspace{10mm}

\begin{abstract}
  
We consider models that generate hierarchies via the separation of fermion
wavefunctions in higher-dimensional spaces.  We calculate the effects of gauge
interactions between fermions and show that these are important and could help
to explain (i) why the heaviest known fermion is a charge 2/3 quark, rather
than a charge $-1/3$ quark or a lepton, (ii) why this fermion has a mass $m_t$
comparable to the electroweak symmetry breaking scale $M_{ew}$, (iii) the
patterns $m_t >> m_b > m_\tau$ and $m_c >> m_s > m_\mu$, and (iv) the smallness
of neutrino masses.

\end{abstract}

\vspace{16mm}

\pagestyle{empty}
\newpage

\pagestyle{plain}
\pagenumbering{arabic}
\renewcommand{\thefootnote}{\arabic{footnote}}
\setcounter{footnote}{0}

The explanation of the spectrum of quark and charged lepton masses and the
Cabibbo-Kobayashi-Maskawa (CKM) quark mixing matrix is an outstanding challenge
that requires physics beyond the standard model (SM).  There are several
general features that one would like to understand.  Why is the heaviest known
fermion a charge 2/3 quark, rather than a charge $-1/3$ quark or a lepton?  Why
does this heaviest known fermion, the top quark, have a mass that is comparable
to the electroweak symmetry breaking (EWSB) scale $M_{ew}=v/\sqrt{2}=174$ GeV,
where $v=2^{-1/4}G_F^{-1/2}=246$ GeV?  Why, in each generation, are the quarks
heavier than the leptons and why, in the heavier two generations, is the mass
of the charge 2/3 quark greater than the mass of the charge $-1/3$ quark? Why
are neutrino masses, for which there is increasingly strong evidence, so much
smaller than the masses of quarks and charged leptons?  Recently, a new
approach to fermion mass hierarchies has been studied, in which one assumes an
underlying higher-dimensional spacetime and obtains the hierarchies from the
localization of fermion wavefunctions at different points in the higher
dimensions \cite{as}-\cite{branco}.  Here we study effects of gauge
interactions between fermions in this type of theory.  We show that these
effects are important and could help to explain the above-mentioned features of
fermion masses. We use the notation $u_i$, $d_i$, and $e_i$ with $i=1,2,3$ to
refer, respectively, to $u,c,t$, $d,s,b$, and $e,\mu,\tau$.

Let us briefly describe the framework.  For each
generation, we denote the left-handed fermion fields as $Q_i$ and $L_i$ for the
quark and lepton SU(2) doublets and $u^c_i$, $d^c_i$, and $e^c_{iL}$ for the
SU(2) singlets.  Usual spacetime coordinates are denoted as $x_\nu$, 
$\nu=0,1,2,3$ and the $n$ extra coordinates as
$y_\lambda$; for definiteness, the latter are taken to be compact.  Generic
fermions fields are denoted $\Psi(x,y)=\psi(x)\chi(y)$.  In the extra
dimensions the gauge, Higgs, and fermion fields are assumed to have support in
an interval $0 \le y_\lambda \le L$.  The $d=4+n$-dimensional fields thus have
Kaluza-Klein (KK) mode decompositions.  We shall work in a low-energy effective
field theory (EFT) approach.  The gauge fields extend over the interval $0 <
y_\lambda < L$, consistent with the observed universality of gauge-fermion
couplings.  A similar assumption is made for the Higgs field(s).  The fermion
wavefunctions are localized at different values of $y_\lambda$. 
This could occur in string theories \cite{orb}, but here, for technical 
simplicity, we assume that it occurs in a field-theoretic
manner.  For example, for the case $n=1$, before the inclusion of gauge 
interactions, consider the action for the quarks:
\beqs 
S & \propto & \int d^4x dy \ \Biggl [ \sum_i \bar
\Psi_{Q_i} (i \gamma \cdot \partial + L^{1/2}\Phi - {\cal M}_{0,Q_i})
\Psi_{Q_i} + \sum_i\sum_{f=u,d}\bar \Psi_{f^c_i} (i \gamma \cdot \partial +
L^{1/2} \Phi - {\cal M}_{0,f_i} ) \Psi_{f^c_i} \cr\cr & &
+ L^{1/2}\sum_{i,j}( \kappa_{d,ij} \Psi_{Q_i}^T C_5 \Psi_{d^c_j} H_d +
\kappa_{u,ij} \Psi_{Q_i}^T C_5 \Psi_{u^c_j} H_u + h.c.) \Biggr ] 
\label{lag}
\eeqs 
where $\gamma \cdot \partial$ and 
$C_5$ are the five-dimensional Dirac operator and charge conjugation matrix.
In (\ref{lag}), in the SM, $H_d=H^\dagger$ and $H_u=\tilde H^\dagger$, with $H$
being the SM Higgs and $\tilde H = i\sigma_2H^*$; more generally, $H_u$ and
$H_d$ may be independent Higgs fields, as in the minimal supersymmetric SM
(MSSM).  A similar formula holds for leptons.  The proportionality factor in
(\ref{lag}) is chosen to yield a canonically normalized 4D action.  If
$\Phi(x,y)$ has the usual kink solution $\Phi=\Phi_0\tanh(\mu y)$, this traps
each fermion to a domain wall at $y_{f_i}=\ell_{f_i}=-{\cal M}_{0,f_i}/\mu^2$
\cite{trap} with localization length $\mu^{-1}$.  Starting with Dirac fermions
in the five-dimensional space, this trapping mechanism yields a chiral theory
in which only left-handed fermions are trapped on the physical 4D domain wall.
In the standard model, the fermions trapped to the physical domain wall are
then $Q_i$, $u^c_i$, $d^c_i$, $L_i$, and $e^c_i$, $i=1,2,3$.  More generally,
one considers the possibility of fermion localization with $n > 1$. 
In the quark sector, for $N_g$ generations, one can 
choose the values of the $n(3N_g-1)$
coordinate differences for the wavefunction centers $\ell_{f_i}$ to fit the
$2N_g$ quark masses and $(N_g-1)^2$ parameters determining the CKM matrix,
i.e. the $N_p=N_g^2+1$ physical parameters. 

If $\mu^{-1} << L$; then, to a good approximation, a
generic fermion wavefunction is a gaussian of the form $\Psi(x,y) =
\psi(x)\chi(y)$ with $\chi(y) \propto e^{-\mu^2(y-\ell)^2/2}$, where 
$(y-\ell)^2 = \sum_{\lambda=1}^n (y_\lambda-\ell_\lambda)^2$.  Performing the
integration over $y$, one obtains the 4D quark Yukawa couplings 
\beq 
S_Y =
\sum_{i,j} \kappa^{(4D)}_{d,ij} \int d^4x \bar \psi_{d_j R}(x) \psi_{Q_i
L}(x)H_d(x) + \sum_{i,j} \kappa^{(4D)}_{u,ij} \int d^4x \bar \psi_{u_j
R}(x) \psi_{Q_i L}(x)\tilde H_u(x) + h.c.
\label{4dlag}
\eeq
where 
\beq
\kappa^{(4D)}_{f,ij}=e^{-\mu^2(\ell_{Q_i}-\ell_{f^c_j})^2/4} \kappa_{f,ij} \ .
\label{ksup}
\eeq
Thus, separations $|\ell_{Q_i}-\ell_{f^c_j}|$ that are moderate in units
of $\mu^{-1}$ produce a strong gaussian suppression of fermion overlaps and
hence of the associated 4D Yukawa couplings.  Similar comments apply for the
charged leptons.  Since the purpose of this type of model is to derive a
hierarchy without putting it in initially, one takes the $\kappa_{f,ij} \sim
O(1)$. By choosing the different separations, one can account for 
the observed fermion mass and mixing angle hierarchies in terms of the
relative locations of fermions in the extra dimensions and further
explain proton longevity and the weakness of flavor-changing neutral current
processes.  This type of model involves three general length scales: $L$,
$\mu^{-1}$, and, since it is a low-energy effective field theory (EFT), a
high-energy cutoff, $\Lambda$. Besides the $\mu^{-1} << L$ condition, one
requires $\mu << \Lambda$ for the self-consistency of the theory.  Possible
values for these parameters, consistent with experimental constraints are sim
$L^{-1} \sim 100$ TeV, $\mu/L^{-1} \simeq 30$, and $\Lambda/\mu \sim 20$
\cite{as}-\cite{greview}, \cite{low} 

We proceed to incorporate the SM SU(3) $\times$ SU(2) $\times$ U(1)$_Y$ gauge
interactions.  Our notation is indicated by the covariant derivative on
quarks, $D_\mu = \partial_\mu - ig_3 C_\mu - ig_2 A_\mu P_L
-i(g^\prime/2)(Y_LP_L+Y_RP_R) B_\mu$, where $C_\mu = \sum_{a=1}^8 C^a_\mu
(\lambda_a/2)$, $A_\mu = \sum_{a=1}^3 A^a_\mu (\tau_a/2)$, $C^a_\mu$,
$A^a_\mu$, and $B_\mu$ are the SU(3) color, SU(2)$_L$, and weak hypercharge
gauge bosons, and $P_{L,R}$ are chiral projection operators. Thus,
$g^\prime/g_2 = \tan\theta_W$ and $g^\prime=(3/5)^{1/2}g_1$, where $g_i$, 
$i=1,2,3$ are the couplings that would unify ($=g$) at a high mass scale in 
a grand unified theory \cite{gu}.

Because the fermions are localized in the higher dimensions, with the
wavefunction factorization given above, they are essentially static as
functions of $y$, so we need only consider the Coulombic interaction between
them.  Since $\mu^{-1} << L$, the Coulomb potential is
that for the full $d=4+n$ dimensional space.  Also $\mu >> L^{-1} >> M_{ew}$,
so (i) the relevant gauge fields are massless on
this scale, and (ii) the color contribution is perturbatively calculable to
good accuracy.  The Yukawa operators $\bar d_{jR} Q_{iL} H_d$ and 
$\bar u_{jR} Q_{iL} H_u$ yield, via the Higgs vevs, the 
bilinears $\bar d_{jR} d_{iL}$ and $\bar u_{jR} u_{iL}$.  Each of these
bilinears involves only the one SU(2) nonsinglet fermion, $Q_{iL}$, so only 
the (vectorial) SU(3)$_c$ color and (chiral) U(1)$_Y$ hypercharge 
interactions between these fermions contribute. 
The distance between the centers of the wavefunctions of the fermions $f_{iL}$
and $f^c_{jL}$ at $(x,\ell_{f_i})$ and $(x,\ell_{f^c_j})$ is
$d_{f,ij}=|\ell_{f_i}-\ell_{f^c_j}|$.  Using $Y_{Q_L}
=1/3$, $Y_{L_L}=-1$, $Y_{f_R}=2Q_f$, the Coulomb interaction energy of $f_{iL}$
and $\bar f_{jR}$ is
\beq
V_{Coul}(y)=-\frac{a_f L^n}{|y|^{1+n}}
\label{vf}
\eeq
with $|y|=d_{f,ij}$, where $a_f = c_f/A_{3+n}$, 
$A_\ell=2\pi^{\ell/2}/\Gamma(\ell/2)$ is the area of the unit sphere $S^\ell$,
and 
\beq
c_u=\frac{4}{3}g_3^2 + \frac{4}{15}g_1^2 \ , 
\label{cu}
\eeq
\beq
c_d=\frac{4}{3}g_3^2 - \frac{2}{15}g_1^2 \ , 
\label{cd}
\eeq
\beq
c_e=\frac{6}{5}g_1^2 \ . 
\label{ce}
\eeq
For example, in the operator $\bar u_{jR} u_{iL}$ one has two fundamental
representations of color SU(3) contracted to a singlet, so the color 
interaction is attractive, and the coupling constant dependence is 
$-(4/3)g_3^2$ \cite{c2}.  Since the hypercharges $Y$ of
the $u_{iL}$ and $\bar u_{jR}$ are 1/3 and $-4/3$, the hypercharge
interaction is $(1/3)(-4/3)(g^\prime)^2= -(4/15)g_1^2$, as given above. 
The normalization in (\ref{vf}) satisfies the requirement
that as $|y|$ increases through the value $L$ beyond which the effective 
dimension of spacetime is 4, the potential matches the usual 4D Coulomb 
potential. Note the general inequality
\beq
c_u > c_d \ge c_e \ . 
\label{ineq}
\eeq
If one makes the additional assumption of gauge coupling unification $g_i=g$ 
at some scale $M_U$ with $\mu > M_U \gsim L^{-1}$, then for
distances $d$ such that $\Lambda^{-1} < d < M_U^{-1}$, $c_u=(8/5)g^2$ and
$c_d=c_e=(6/5)g^2$. 

Before inclusion of gauge interactions, the suppression in (\ref{ksup}) can be
treated as the result of the tunnelling of each fermion to the midpoint of
the line joining $\ell_{f_i}$ and $\ell_{f^c_j}$, located at a distance
$d_{f,ij}/2$ from each \cite{as}.  The WKB tunnelling amplitude \cite{wkb} for
each fermion is $\exp(-\int_0^{d_{f,ij}/2} V(r)dr) = \exp(-(\mu
d_{f,ij})^2/8)$, yielding, for the total suppression, the factor $\exp(-(\mu
d_{f,ij})^2/4)$, in agreement with (\ref{ksup}).

We now consider the effect of the Coulomb interactions.  Since $L/d_{f,ij}$
values are reasonably large, the effects of the boundary conditions on the 
gauge fields, e.g., image charges, are small, and we neglect them.  
Given that the fermions have the same
localization lengths, the tunnelling can be
treated in a symmetric way for each pair.  The full potential energy for
$f_i$ at a distance $r$ outward from its wavefunction center toward $f^c_j$ is
$V_f(r)=V_{trap}(r)+V_{Coul}(d_{f,ij}-r)$, where $V_{trap}(r)=\mu^2r$. The 
resultant restoring force
in the ${\hat {\bf r}}$ direction is $F=-\mu^2
+a_f(n+1)L^n/(d_{f,ij}-r)^{n+1}$.  The most dramatic effect occurs if
$F(r=0^+)=-\mu^2 + a_f(n+1)L^n/(d_{f,ij})^{n+2}$ is positive, i.e.,
\beq
(\mu d_{f,ij})^2 < a_f(n+1)\biggl (\frac{L}{d_{f,ij}}\biggr )^n \ ; 
\label{collapse}
\eeq 
here, the Coulomb attraction overwhelms the trapping potential and causes $f_i$
and $f^c_j$ to have wavefunctions that are centered at the same location, at
least to within the distance $\Lambda^{-1}$ down to which the EFT applies.
Hence in this case there is no suppression of the Yukawa coupling
$\kappa^{(4D)}_{f,ij}$, so that, if the higher-dimensional coupling
$\kappa_{f,ij} \sim O(1)$, then the resultant fermion mass is of order
$M_{ew}$.  Let us first concentrate on the third generation and neglect small
mixing effects.  We further focus on the $n=2$ case since a successful fit was
found for this case in \cite{ms} (before the inclusion of gauge interactions).
This fit yielded $\mu d_{u,33}=0.9005$, $\mu d_{d,33}=3.00$, 
and $\mu d_{e,33}=3.15$, with $\mu L \simeq 18$.  We use the illustrative
gauge-coupling unification value $g_i=g$ with $g^2/(4\pi) \simeq 0.04$, so that
$a_u=0.030$ and $a_d=a_e=0.023$.  
Then $(\mu d_{u,33})^2=0.81$, which is smaller than the RHS
of (\ref{collapse}), viz., 46, while $(\mu d_{f,33})^2 >$ RHS(\ref{collapse}) 
for $f=d,e$. Thus, 
with these input values, the Coulomb interactions cause $t_L$ and $t^c_L$
wavefunctions to be centered essentially on top of each other, but do not
overwhelm the trapping of other fermions to their domain walls.  Generalizing,
we can say that if a fermion has $\mu d_{f,33} \lsim O(1)$, i.e., is moderately
heavy, then, for a value of $\mu L \sim 20$ that is phenomenologically
acceptable, LHS(\ref{collapse}) $<$ RHS(\ref{collapse}), so that the Coulomb
interaction can dominate over the trapping interaction with $\Phi$ and cause 
the chiral components of this fermion to lie essentially on top of each other,
leading to $m_f \sim M_{ew}$ if $\kappa_{f,33} \sim O(1)$.  For plausible input
values, this can happen for a charge 2/3 quarks while the charge $-1/3$ quarks
and leptons remain trapped on their domain walls.  This could thus help to 
explain the fact that the heaviest known fermion, and the only fermion
with a mass comparable to the EWSB scale $M_{ew}$, is a charge 2/3 quark,
rather than a charge $-1/3$ quark or a charged lepton and thus could provide
deeper insight into properties (i) and (ii) in the abstract.  

Although we have discussed this Coulomb-induced collapse in the model of
\cite{as}, we also note that, more generally, gauge interactions could also be
relevant to models with dynamical EWSB involving multifermion operators in
which a $\langle \bar f f \rangle$ condensate forms.  We suggest that these
gauge interactions and the inequality (\ref{ineq}) could explain why in such
dynamical EWSB models it is the $\langle \bar t t \rangle$ condensate that
forms rather than a $\langle \bar b b \rangle$ or $\langle \bar \tau \tau
\rangle$ condensate.

Just as a WKB approximation can be used to infer the result (\ref{ksup}) before
inclusion of the Coulomb effects, so also it can be used to calculate the
latter effects.  We next do this for fermions for which 
this attraction does not overwhelm the trapping to domain walls. 
Consider the symmetric path where $f_i$ and $f^c_j$ each tunnel a distance $r$ 
toward each other from their respective centers $(x,\ell_{f_i})$ and 
$(x,\ell_{f^c_j})$ so that they are a distance $2\epsilon_{f,ij}=d_{f,ij}-2r$ 
apart.  The classical turning points $(r_t)_{f,ij}$ 
occur where the total potential energy 
\beq
V_{tot}(r) = 2\mu^2 r - \frac{a_f L^n}{(d_{f,ij}-2r)^{n+1}}
\label{vrtot}
\eeq
vanishes. We have $\epsilon_{f,ij}=(1/2)d_{f,ij}-(r_t)_{f,ij}$ and define 
$\eta_{f,ij}=2 \epsilon_{f,ij}/d_{f,ij}$. 
Then the equation for the turning point becomes 
\beq
\eta_{f,ij}^{n+1}(1-\eta_{f,ij})=b_f
\label{eta_eq}
\eeq
where $b_f=a_f(\mu d_{f,ij})^{-2}(L/d_{f,ij})^n$.  Since the mutual tunnelling
by the fermions does not have to proceed further than to the midpoint between
them, we are only interested in the range $0 \le \eta_{f,ij} \le 1$.  For 
this range, the LHS of (\ref{eta_eq}) increases from 0 to a maximum value 
$b_{f,m}=(n+1)^{n+1}/(n+2)^{n+2}$ at $(\eta_{f,ij})_m=(n+1)/(n+2)$ 
and then decreases to 0 again at $\eta_{f,ij}=1$.  If $b_f \le b_{f,m}$, 
(\ref{eta_eq}) has physical solutions.  For $b_f = b_{f,m}$, there is a 
unique physical solution, $\eta_{f,ij}=(\eta_{f,ij})_m$.  
For $0 < b_f < b_{f,max}$, the two turning points are given by 
$r^{(1,2)}_{f,ij}=(1-\eta^{(2,1)}_{f,ij})d_{f,ij}/2$ with 
$0 \le \eta^{(1)}_{f,ij} < (n+1)/(n+2) < \eta^{(2)}_{f,ij} < 1$. 
The relation between the higher-dimensional and 4D Yukawa couplings is then
given by 
\beq
\kappa^{(4D)}_{f,ij} = w_{f,ij} \ \kappa_{f,ij}
\label{kapcoul}
\eeq 
where the overlap factor is, in the WKB approximation, 
$w_{f,ij}=\exp(-J_{f,ij})$, with
\beqs
& & J_{f,ij} = \int_{r^{(1)}_{f,ij}}^{r^{(2)}_{f,ij}} V_{tot}(r)dr \cr\cr
& & = (\eta^{(2)}_{f,ij}-\eta^{(1)}_{f,ij})\Biggl [ 
\biggl ( \frac{\mu d_{f,ij}}{2}\biggr )^2 [ 2 -(\eta^{(1)}_{f,ij}+
\eta^{(2)}_{f,ij})] - \frac{a_f}{2 n \eta^{(1)}_{f,ij}\eta^{(2)}_{f,ij}}
\Bigl (\frac{L}{d_{f,ij}}\Bigr )^n \Biggr ] \ . 
\label{expint}
\eeqs 
The integral extends over the classically forbidden region between the
two turning points \cite{wkb}.  For the $n=2$ parameters above, we find that 
the gauge interaction increases the overlap factors $w_{d,33}$ and 
$w_{e,33}$ relevant for $m_b$ and $m_\tau$ by 22 \% from 0.77 to 0.98 
and from 0.76 to 0.93, respectively.  Thus, gauge interactions have a 
significant enhancement effect on the wavefunction overlaps and
hence Yukawa couplings.  Although we have concentrated on the case $g_3=g_1$,
the more general case $g_3 > g_1$ would lead to further enhancement of the
masses of the charge 2/3 and $-1/3$ quarks relative to those of the charged
leptons.  Of course, this is not an {\it ab initio} calculation of the fermion
mass spectrum, since it depends on initial inputs for the relative distances
$d_{f,ij}$.  What our calculations show is that gauge interactions,
together with the inequality (\ref{ineq}), could help to explain why the quarks
of a given generation are heavier than the charged lepton and why, at least
for the higher two generations, $m_{Q=2/3} > m_{Q=-1/3}$, i.e. property (iii).

For the first generation, given the smallness of $m_u$ and $m_d$ \cite{qmass},
a model for these masses should take account of the off-diagonal quantities
$\kappa^{(4D)}_{f,ij}$, $ij=12,21$.  Indeed, if the $N_g=1,2$ subsectors of 
the mass matrices for the charge 2/3 and $-1/3$ quarks have the form (after
allowed rephasings so that $A^{(f)}_{22}$ is real and positive) 
\beqs
M^{(f)}=\pmatrix{\sim 0 & A^{(f)}_{12} \cr 
A^{(f)}_{21} & A^{(f)}_{22}}
\eeqs
where $|A^{(f)}_{ij}|/A^{(f)}_{22} << 1$ for $ij=12,21$ and $\sim 0$ 
means a negligibly small entry, the eigenvalues 
have the form $\lambda^{(f)}_2 \simeq A^{(f)}_{22}$ and 
$|\lambda^{(f)}_1| \simeq |A^{(f)}_{12}A^{(f)}_{21}|/A^{(f)}_{22}$ (with 
$\lambda^{(u)}_2=m_c$, $\lambda^{(d)}_2=m_s$, $|\lambda^{(u)}_1|=m_u$,
$|\lambda^{(d)}_1|=m_d$).  Since 
$A^{(u)}_{22} \simeq m_c >> A^{(d)}_{22} \simeq m_s$ and since these enter in 
the denominators of the expressions for the lighter eigenvalues, if 
$A^{(u)}_{ij}$ and $A^{(d)}_{ij}$, $ij=12,21$ are not too different, this 
seesaw effect could accomodate the fact that $m_u < m_d$. That is, these 
lightest quark masses could arise primarily by mixing, and hence could avoid 
the generic pattern $m_{u_i} > m_{d_i} > m_{e_i}$ for the heavier two 
generations $i=2,3$. 

An important comment concerns the calculability of these gauge interaction
effects.  The fact that fermions with stronger gauge interactions (color and
$U(1)_Y$ or $U(1)_{em}$) are more massive is very suggestive.  Yet conventional
attempts to explain this via radiative corrections in usual quantum field
theory encounter the obstacle that perturbative gauge boson couplings preserve
chirality and cannot generate masses from originally massless fermions. Once
fermion masses are generated, by the Higgs mechanism or in some other way,
radiative gauge boson corrections modify the tree-level masses; however, these
corrections are divergent, so that the physical masses are arbitrary numbers
which are inserted to fit experiment.  In contrast, our effects are calculable
and finite.  Even in the case where they overwhelm the domain-wall trapping,
and hence there can be sensitivity to $\Lambda$, this just results in 
$w_{f,ij} \simeq 1$.

We next consider neutrinos. In the minimal approach, where one avoids
electroweak-singlet neutrinos, the Majorana mass term 
$\nu_{iL}^T C (M_L)_{ij} \nu_{jL}$ would arise from the operator 
\beq 
{\cal O} = \frac{1}{M_X}\sum_{i,j}h_{ij}
(\epsilon_{ap}\epsilon_{bq}+\epsilon_{aq}\epsilon_{bp}) L^{T a}_i C_d L^b_j H^p
H^q + h.c. 
\eeq 
(where $a,b,p,q$ are SU(2) indices) via the vevs of the Higgs.  Here, $M_X$ 
could be $\sim O(\Lambda)$. Since $L^a_i$ and $L^b_j$ are coupled
to $I=1$ and since they have the same hypercharge ($Y=-1$), both the SU(2) 
and U(1)$_Y$ Coulomb interactions are repulsive for the $L^a_i$, 
$L^b_j$ pair, separated by the distance $d_{LL,ij}$.  This would tend to move
the corresponding wavefunctions farther apart and thereby suppress the
wavefunction overlap, naturally leading to small 4D operator coefficients.  For
the case $i=j$, the Coulomb self-energy is especially large, $\sim g_k^2
\Lambda$, $k=1,2$.  This would strongly suppress the diagonal contributions to
the left-handed neutrino Majorana mass matrix.  In turn, this could naturally
lead to large mixing.  Given that the leptonic mixing matrix 
$U=U_{\nu}U_e^\dagger$, if $U_e$ has only small mixing, a (2,3) 
subsector of the $\nu_{iL}$ Majorana mass matrix $M_L$ (which always 
satisfies $(M_L)_{ij}=(M_L)_{ji}$) of the form 
\beqs
(M_L)=\pmatrix{\sim 0 & C \cr
C & \sim 0}
\eeqs
will yield maximal $\nu_\mu,\nu_\tau$ mixing, in agreement with the 
atmospheric neutrino data \cite{atm,nur}. 

Some of the features that we have found may well transcend the specific EFT
approach used here.  Indeed, the latter leaves a number of open questions.  Why
are there three generations of standard model fermions standard model with
light neutrinos?  What are the implications of the fact that the theory is
nonrenormalizable in $d > 4$ dimensions?  What further insights can these
models give to EWSB (e.g. \cite{dob}) ?  What physics gives rise to the
restriction of the gauge and matter fields to $M \times [0,L]^n$, where $M$ is
Minkowski space, and to the localization of the fermion wavefunctions?  The
latter problem may well be related to the open string/brane construction of
gauge theories.  We speculate that the localization of fermions and resultant
gaussian profile might be achieved without introducing the $\Phi$ field and
higher-dimensional Yukawa couplings and masses and will return to this issue
later, together with further study of quark and lepton masses and mixing.

In summary, we have shown that the inclusion of gauge interactions in models
with fermion wavefunctions separated in extra dimensions has important
consequences and could help to provide an explanation of several of the most
basic features of the known fermion masses.  This explanation is pleasingly
minimal in that it makes use only of the established SM gauge transformation
properties of the known fermions, albeit in a new context.

S. N. would like to thank the Israeli Academy for Fundamental Research for a
grant.  The research of R. S. was partially supported by the NSF grant
PHY-97-22101.  S. N. thanks the Yang ITP for hospitality in Sept.-Oct. 2000,
and R. S. thanks Tel Aviv University for hospitality during visits when parts
of this research were carried out.

\vfill
\eject


\begin{thebibliography}{99}
\bibitem{gu}{Rapid power law running of gauge and Yukawa couplings occurs in
the present models since $d > 4$ \cite{low}.  This can lead to precocious
gauge coupling unification.  However, the placement of fermions at
different positions explicitly breaks both SU(5) and SO(10); e.g., as 
components of the $10_L$ of SU(5), $Q_i$, $u^c_i$, and $e^c_i$ would
necessarily have wavefunctions centered at the same point.}
\bibitem{c2}{For general $N_c$, the factor 4/3 is 
$C_2(fund.)=(N_c^2-1)/(2N_c)$.}
\bibitem{qmass}{We refer to the running quark masses evaluated at a common 
scale $>> \Lambda_{QCD}$.}
\end{thebibliography}
\end{document}